\documentclass[seceq]{jpsj2}
\title{Directional Dichroism of X-Ray Absorption in a Polar Ferrimagnet GaFeO$_3$}

\author{Jun-ichi \textsc{Igarashi}$^{1}$\thanks{E-mail:jigarash@mx.ibaraki.ac.jp} 
and Tatsuya \textsc{Nagao}$^{2}$}

\inst{$^{1}$Faculty of Science, Ibaraki University, Mito, Ibaraki 310-8512, Japan \\
$^{2}$Faculty of Engineering, Gunma University, Kiryu, Gunma 376-8515, Japan}

\abst{We study the directional dichroic absorption spectra in the x-ray region
in a polar ferrimagnet GaFeO$_3$. 
The directional dichroism on the absorption spectra at the Fe pre-$K$-edge
arises from the $E1$-$E2$ interference process through the hybridization 
between the $4p$ and $3d$ states in the noncentrosymmetric  environment of 
Fe atoms. We perform a microscopic calculation of the spectra on a model 
of FeO$_6$ with reasonable parameter values for Coulomb interaction and
hybridizations. We obtain the difference in the absorption coefficients
when the magnetic field is applied parallel and antiparallel to the $c$ axis.
The spectra thus obtained have similar shapes to the experimental curves 
as a function of photon energy in the Fe pre-$K$-edge region,
although they have opposite signs.}

\kword{directional dichroism, magnetoelectric effect, GaFeO$_{3}$, noncentrosymmetricity,
$E$1-$E$2 interference, x-ray $K$-edge absorption}

\begin{document}
\maketitle

\section{Introduction}
Gallium ferrate (GaFeO$_3$) exhibits simultaneously spontaneous electric 
polarization and magnetization at low temperatures. 
This compound was first synthesized by Remeika,\cite{Remeika1960} 
and the large magnetoelectric effect was observed by Rado.\cite{Rado1964}
Recently, untwinned large single crystals have been prepared,\cite{Arima2004} 
and the optical and x-ray absorption measurements have been carried out 
with changing the direction of magnetization.\cite{Jung2004,Kubota2004}
The purpose of this paper is to analyze the magnetoelectric effects on
the x-ray absorption spectra and to elucidate the microscopic origin 
by carrying out a microscopic calculation of the spectra.

The crystal of GaFeO$_3$ has an orthorhombic unit cell with the space group
$Pc2_{1}n$.\cite{Wood1960}
Each Fe atom is octahedrally surrounded by O atoms.
With neglecting slight distortions of octahedrons, Fe atoms are regarded as 
slightly displaced from the center of the octahedron;
the shift is $0.26 \textrm{\AA}$ at Fe1 sites 
and $-0.11 \textrm{\AA}$ at Fe2 sites along the $b$ axis.\cite{Arima2004} 
Thereby the spontaneous electric polarization is generated along the $b$ axis. 
Note that two kinds of FeO$_6$ clusters exist for both Fe1 and Fe2 sites, 
one of which is given by rotating the other by an angle $\pi$ 
around the $b$ axis.
As regards magnetic properties, the compound behaves as a ferrimagnet,
\cite{Frankel1965,Arima2004} with the local magnetic moments at Fe1 and Fe2 
sites aligning antiferromagnetically along the $c$ axis. 
One reason for the ferrimagnetism may be that the Fe occupation at Fe1 and 
Fe2 sites are slightly different from each other.\cite{Arima2004} 
In the present analysis,
we neglect such a small deviation from a perfect antiferromagnet.

\begin{figure}[tb]
\begin{center}\includegraphics[width=8.0cm]{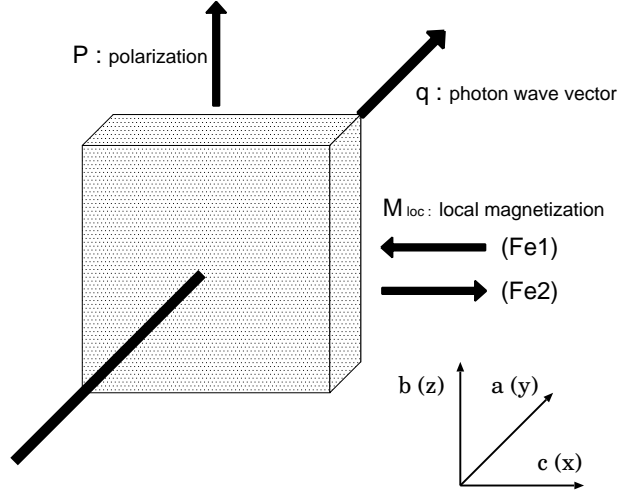}
\caption{Geometry of absorption experiment.\cite{Kubota2004} 
X-rays propagate along the $a$ axis
with polarization along the $b$ axis or the $c$ axis.
The electric dipole moment is along the $b$ axis. 
When the magnetic field is applied to the positive direction of the $c$ axis,
the sublattice magnetization is directed to the negative and positive 
directions of the $c$ axis at Fe1 sites and at Fe2 sites, respectively. 
When the magnetic field is reversed, the sublattice magnetization
is reversed.
\label{fig.setup}}
\end{center}
\end{figure}

In the $K$-edge absorption experiment,\cite{Kubota2004}
the x-rays propagated along the
positive direction of the $a$ axis, and the magnetic field was applied 
along the $\pm c$ axis, as illustrated in Fig.~\ref{fig.setup}.
The difference of the absorption coefficient was measured between the two 
directions of magnetic field. 
As will be shown later [eq. (\ref{eq.toroidal})], these difference spectra have
characteristic dependence on the polarization and magnetization,
and would be termed as \emph{magnetoelectric} spectra.\cite{Kubota2004}
Since the compound is a ferrimanget, reversing the direction of applied magnetic 
field results in reversing the direction of the local magnetic moment of Fe
atoms. 

In the analysis of absorption spectra, we consider only the processes on 
Fe atoms, since the $1s$-core state is well localized on Fe atoms.
In addition to the $E1$-$E1$ and $E2$-$E2$ processes, we formulate 
the $E1$-$E2$ interference process which gives rise to the magnetoelectric
spectra. The contribution of this process  arises from
the mixing of the $3d^{6}\underline{1s}$-configuration 
to the $4p3d^5\underline{1s}$-configuration, where $\underline{1s}$ 
indicates the presence of a $1s$-core hole. Such mixings exist only under
the noncentrosymmetric environment.
In the present analysis, we describe the $E1$-$E2$ process by employing
a cluster model of FeO$_{6}$, where all the $3d$ and $4p$ orbitals of Fe 
atoms and the 2p orbitals of O atoms as well as the Coulomb and 
the spin-orbit interactions in the $3d$ orbitals are taken into account. 
We obtain an effective hybridization between the $4p$ and $3d$ states 
in addition to the ligand field on the $3d$ states through the hybridization 
with the O $2p$ states. On the basis of these frameworks, 
we clarify various symmetry relations 
to the $E1$-$E2$ process and the relation between the nonreciprocal
directional dichroism and the anapole moment. Furthermore, we numerically 
calculate the absorption spectra as a function of photon energy
by diagonalizing the Hamiltonian matrix in the $3d^6$- and 
$4p3d^5$-configurations.
The shapes of magnetoelectric spectra as a function of photon energy
are found similar to the experimental curves
but their signs are opposite to the experiment
in the pre-$K$-edge region.\cite{Kubota2004} 
The origin for the opposite sign is not known.
Note that the magnetoelectric spectra in the optical absorption have been
obtained in agreement with the experiment by using 
the same cluster model.\cite{Igarashi2009} 

This paper is organized as follows. In \S 2, we introduce a cluster model
of FeO$_6$. In \S 3, we describe the x-ray transition operators
associated with Fe atoms.
In \S 4, we derive the formulas of x-ray absorption, and
present the calculated spectra in comparison with the experiment.
The last section is devoted to concluding remarks.

\section{Electronic Structures}

\subsection{F\lowercase{e}O$_6$ cluster and $4p$ band\label{sect.2.1}}

In a FeO$_6$ cluster, we consider the $1s$, $3d$ and $4p$ states 
in the Fe atom, and the $2p$ states in O atoms.
The Hamiltonian may be written as
\begin{eqnarray}
 H &=& H^{3d} + H^{2p} + H_{\rm hyb}^{3d-2p} + H^{4p} 
 + H_{\rm hyb}^{4p-2p} \nonumber \\
&+& H^{1s} + H^{1s-3d} + H^{1s-4p}, \label{eq.Ham}
\end{eqnarray}
where 
\begin{eqnarray}
 H^{3d} & = & \sum_{m\sigma}E_{m}^{d}
d^{\dagger}_{m\sigma}d_{m\sigma} \nonumber \\
  &+& \frac{1}{2}\sum_{\nu_{1}\nu_{2}\nu_{3}\nu_{4}}
  g\left(\nu_{1}\nu_{2};\nu_{3}\nu_{4}\right)d_{\nu_{1}}^{\dagger}
  d_{\nu_{2}}^{\dagger}d_{\nu_{4}}d_{\nu_{3}} \nonumber\\
& + & \zeta_{3d}\sum_{mm'\sigma\sigma'}
 \langle m\sigma|{\bf L}\cdot{\bf S}|m'\sigma'\rangle
  d^{\dagger}_{m\sigma}d_{m'\sigma'} \nonumber \\
 &+&{\bf H}_{\rm xc}\cdot \sum_{m\sigma\sigma'}
   ({\bf S})_{\sigma\sigma'}d^{\dagger}_{m\sigma}d_{m\sigma'}, 
\label{eq.H3d}\\
 H^{2p} &=& \sum_{j\eta\sigma}E^{p}p^{\dagger}_{j\eta\sigma}p_{j\eta\sigma},\\
 H_{\rm hyb}^{3d-2p} &=& \sum_{j\eta\sigma m}t_{m\eta}^{3d-2p}(j)
   d_{m\sigma}^{\dagger}p_{j\eta\sigma}+{\rm H.c.}, \\
 H^{4p} &=& \sum_{{\bf k}\eta'\sigma}\epsilon_{4p}({\bf k})
   p'^{\dagger}_{{\bf k}\eta'\sigma} p'_{{\bf k}\eta'\sigma},\\
 H_{\rm hyb}^{4p-2p} &=& \sum_{j\eta\sigma\eta'} 
 t_{\eta'\eta}^{4p-2p}(j)
   p'^{\dagger}_{\eta'\sigma} p_{j\eta\sigma}+{\rm H.c.}, \\
 H^{1s} &=& \epsilon_{1s}\sum_{\sigma} s^{\dagger}_{\sigma}s_{\sigma}, \\
 H^{1s-3d} &=& U^{1s-3d}\sum_{m\sigma\sigma'}
      d^{\dagger}_{m\sigma}d_{m\sigma}s^{\dagger}_{\sigma'}s_{\sigma'}, \\
 H^{1s-4p} &=& U^{1s-4p}\sum_{\eta'\sigma\sigma'}
  p'^{\dagger}_{\eta'\sigma}p'_{\eta'\sigma}s^{\dagger}_{\sigma'}s_{\sigma'}.
\end{eqnarray}

The energy of $3d$ electrons can be described by
$H^{3d}$ [eq.~(\ref{eq.H3d})] where
$d_{m\sigma}$ represents an annihilation operator of a $3d$ electron 
with spin $\sigma$ and orbital $m$ ($=x^2-y^2,3z^2-r^2,yz,zx,xy$).
The symbol $E_m^d$ refers to the energy of $3d$ state with orbital $m$. 
The second and third terms of eq.~(\ref{eq.H3d}) represent the 
intra-atomic Coulomb and spin-orbit interactions for $3d$ electrons,
respectively.
The matrix elements 
$g\left(\nu_{1}\nu_{2};\nu_{3}\nu_{4}\right)$ are expressed in terms of the
Slater integrals $F^{0}$, 
$F^{2}$, and $F^{4}$ [$\nu$ stands for $\left(m,\sigma\right)$], and
the spin-orbit coupling is $\zeta_{3d}$.
We evaluate atomic values of $F^2$, $F^4$ and $\zeta_{3d}$ 
within the Hartree-Fock (HF) approximation,\cite{Cowan1981} and multiply 
$0.8$ to these atomic values in order to take account of the slight
screening effect. On the other hand, we multiply 0.25 to the atomic value 
for $F^0$, since $F^{0}$ is known to be considerably screened 
by solid-state effects.  
The last term in eq.~(\ref{eq.H3d}) describes the energy arising from 
the exchange interaction with neighboring Fe atoms,
where $({\bf S})_{\sigma\sigma'}$ represents the matrix element of
the spin operator of $3d$ electrons. 
The exchange field ${\bf H}_{\rm xc}$ has a dimension of energy,
and is $\sim k_{\rm B}T_{c}/4$ with $T_{c}\sim 250$ K. Note that this term
has merely a role of selecting the ground state by lifting the degeneracy 
and therefore the spectra depend little on its absolute value. 
The ${\bf H}_{\rm xc}$ is directed to the negative direction of the $c$ axis 
at Fe1 sites when the magnetic field is applied along the positive 
direction of the $c$ axis.

The energy of oxygen $2p$ electrons
is given by $H^{2p}$ where 
$p_{j\eta\sigma}$ is the annihilation operator of the $2p$ state of 
energy $E^p$ with $\eta=x,y,z$ and spin $\sigma$ at the oxygen site $j$.
The Coulomb interaction is neglected in oxygen 2p states. 
The $H^{3d-2p}_{\rm hyb}$ denotes the hybridization energy 
between the $3d$ and $2p$ states with the coupling constant $t_{m \eta}^{3d-2p}$. 
The energy of the $2p$ level relative to the $3d$ levels is determined from 
the charge-transfer energy $\Delta$ defined by 
$\Delta=E^{d}-E^{p}+15U(3d^6)-10U(3d^5)$ with $E^d$ being 
an average of $E_m^d$.
The multiplet-averaged $d$-$d$ 
Coulomb interaction in the $3d^6$ and $3d^5$ configurations
are referred to as $U(3d^6)$ and $U(3d^5)$, respectively,
which are defined by $U=F^{0}-\left(2/63\right)F^{2}-\left(2/63\right)F^{4}$.

The $H^{4p}$ represents the energy of the $4p$ states, 
where $p'_{{\bf k}\eta'\sigma}$ is the annihilation operator of the $4p$ 
state with momentum ${\bf k}$, $\eta'=x,y,z$, and spin $\sigma$.
The $4p$ states form an energy band $\epsilon_{4p}({\bf k})$.
The density of states (DOS) of the $4p$ band is inferred from
the $K$-edge absorption spectra\cite{Kubota2004} (see Fig.~\ref{fig.average}).
The $H^{4p-2p}_{\rm hyb}$ represents the hybridization between the $4p$
and oxygen $2p$ states with the coupling constant $t_{\eta' \eta}^{4p-2p}$.
The annihilation operator of the local $4p$
orbital $p'_{\eta'\sigma}$ may be expressed as 
$p'_{\eta'\sigma}=(1/\sqrt{N_0})\sum_{\bf k} p'_{{\bf k}\eta'\sigma}$
($N_0$ is the discretized number of ${\bf k}$-points).

The energy of the $1s$ state is denoted as $H^{1s}$
where $s_{\sigma}$ represents the annihilation operator of the $1s$ state 
with spin $\sigma$.
Finally, the interaction between the $1s$ and 
$3d$ states and that between the $1s$ and $4p$ states
are denoted as $H^{1s-3d}$ and $H^{1s-4p}$, respectively.

Table \ref{table.1} lists the parameter values used in this paper,
which are used in our previous paper analyzing the optical spectra
in GaFeO$_3$,\cite{Igarashi2009} and are consistent with the values 
in previous calculations for Fe$_3$O$_4$.\cite{Chen2004,Igarashi2008}

\begin{table}[tb]
\caption{\label{table.1}
Parameter values for a FeO$_{6}$ cluster in the $3d^5$ configuration,
in units of eV. The Slater-Koster two-center integrals are defined
for the Fe atom at the center of the octahedron.}
\begin{tabular}{lrlr}
\hline
$F^0(3d,3d)$ & 6.39 & $(pd\sigma)_{2p,3d}$ & -1.9 \\
$F^2(3d,3d)$ & 9.64 & $(pd\pi)_{2p,3d}$ & 0.82 \\
$F^4(3d,3d)$ & 6.03 & $(pp\sigma)_{2p,4p}$ & 3.5 \\
$\zeta_{3d}$ & 0.059 & $(pp\pi)_{2p,4p}$ & -1.0 \\
$\Delta$ & 3.3 &  &  \\
\hline
\end{tabular}
\end{table} 

\subsection{Ligand field and effective hybridization between the $4p$ and 
$3d$ states}

Fe atoms are assumed to be displaced from the center of the octahedron
along the $b$ axis. The shift $\delta$ is $0.26 \textrm{\AA}$ at Fe1 sites 
and $-0.11 \textrm{\AA}$ at Fe2 sites.
We evaluate the hybridization matrices $t^{3d-2p}_{m\eta}(j)$ and 
$t^{4p-2p}_{\eta'\eta}(j)$ for the Fe atom at the off-center positions 
by modifying the Slater-Koster two-center integrals for the Fe atom 
at the central position of the octahedron (Table \ref{table.1})
with the assumption that 
$(pd\sigma)_{2p,3d}$, $(pd\pi)_{2p,3d}\propto d^{-4}$,
and $(pp\sigma)_{4p,2p}$, $(pp\pi)_{4p,2p}\propto d^{-2}$ for $d$ being 
the Fe-O distance.\cite{Harrison2004}
Thereby the ligand field Hamiltonian on the $3d$ states is given in
second-order perturbation theory,
\begin{equation}
 \tilde{H}^{3d-3d}= \sum_{mm'\sigma} \tilde{t}_{mm'}^{3d-3d} 
  d_{m\sigma}^{\dagger}d_{m'\sigma} + {\rm H.c.},
\end{equation}
with
\begin{equation}
 \tilde{t}^{3d-3d}_{mm'} = \sum_{j\eta} t^{3d-2p}_{m\eta}(j)
 t^{3d-2p}_{m'\eta}(j)/\Delta,
\end{equation}
where the sum over $j$ is taken on neighboring O sites, and
$\Delta=3.3$ eV is the charge transfer energy defined in \S \ref{sect.2.1}. 
In addition to the ligand field corresponding to the cubic 
symmetry, we have a field proportional to $\delta^2$.
The latter causes extra splittings of the $3d$ levels. 
Similarly, we can
evaluate the effective hybridization between 
the $4p$ and $3d$ states in the form:
\begin{equation}
 \tilde{H}^{4p-3d}= \sum_{\eta'm\sigma} \tilde{t}_{\eta'm}^{4p-3d} 
  p'^{\dagger}_{\eta'\sigma}d_{m\sigma} + {\rm H.c.}
\end{equation}
Here, the effective coupling is defined by
\begin{equation}
 \tilde{t}^{4p-3d}_{\eta'm} = \sum_{j\eta} t^{4p-2p}_{\eta'\eta}(j)
       t^{3d-2p}_{m\eta}(j)/(E^{4p}-E^{2p}),
\end{equation}
where $E^{4p}$ is the average of the $4p$-band energy, which
is estimated as $E^{4p}-E^{2p}\approx 17$ eV. 
The value of coupling coefficient $\tilde{t}^{4p-3d}_{\eta'm}$ is 
nearly proportional to the 
shift $\delta$ of the Fe atom from the center of the octahedron.

\subsection{Ground state}

We assume that Fe ions are in the $d^5$-configuration in the ground state,
which will be denoted as $|\Phi_g(d^5)\rangle$ with eigenenergy $E_g(d^5)$.
We calculate the state by diagonalizing the Hamiltonian 
$H_{3d}+\tilde{H}^{3d-3d}$, where the exchange field ${\bf H}_{\textrm{xc}}$ 
and the displacement of Fe atoms from the center of the octahadron are
different from Fe1 and Fe2 sites. The lowest energy state is characterized as
$^{6}A_{1}$ under the trigonal crystal field when the exchange 
field and the spin-orbit interaction are disregarded. The inclusion of these interactions 
could induce the orbital moment $\hbar\langle L_x\rangle$, but its absolute 
value is given less than $0.004\hbar$. 

\section{Absorption process on F\lowercase{e}}

The interaction between the electromagnetic wave and electrons is described by
\begin{equation}
 H_{\rm int} = -\frac{1}{c}\int {\bf j}({\bf r})\cdot{\bf A}({\bf r})
 {\rm d}^3{\bf r},
\label{eq.int}
\end{equation}
where $c$ stands for the speed of light and ${\bf j}$ represents the current-density operator.
The electromagnetic field ${\bf A}({\bf r})$ for linear polarization
is defined as 
\begin{equation}
 {\bf A}({\bf r}) = \sum_{\bf q}
 \sqrt{\frac{2\pi\hbar^2 c^2}{V\hbar\omega_{\bf q}}}
  {\bf e}c_{\bf q}{\rm e}^{i \textbf{q}\cdot\textbf{r}} 
  + {\rm H.c.},
\end{equation}
where $c_{\bf q}$ and $\hbar\omega_{\bf q}$ are
the annihilation operator and the energy of photon, respectively.
The unit vector of polarization is described by ${\bf e}$. 
We approximate this expression into a sum of the contributions 
from each Fe atom:
\begin{equation}
 H_{\rm int} = -\frac{1}{c}\sum_{{\bf q},i}{\bf j}({\bf q},i)\cdot
 {\bf A}({\bf q},{\bf e},i) + {\rm H.c.},
\end{equation}
with
\begin{eqnarray}
 {\bf j}({\bf q},i) &=& \sum_{nn'} 
  \left[\int {\rm e}^{i{\bf q}\cdot({\bf r}-{\bf r}_i)}
  {\bf j}_{nn'}({\bf r}-{\bf r}_i){\rm d}^3({\bf r}-{\bf r}_i)\right]
    a_{n}^{\dagger}(i)a_{n'}(i) , 
\label{eq.localcurrent}\\
 {\bf A}({\bf q},{\bf e},i) &=& 
 \sqrt{\frac{2\pi\hbar c^2}{V\omega_{\bf q}}}
   {\bf e}c_{\bf q}{\rm e}^{i \textbf{q}\cdot\textbf{r}_i}.
\end{eqnarray}
The local current operator may be described by\cite{Landau}
\begin{eqnarray}
 {\bf j}_{nn'}({\bf r}-{\bf r}_i) &=& \frac{ie\hbar}{2m}
 \big[(\nabla \phi^{*}_n)\phi_{n'}
  - \phi_n^{*}\nabla\phi_{n'}\big] 
  \nonumber\\
  &-& \frac{e^2}{mc}{\bf A}\phi_n^{*}\phi_{n'} 
+\frac{e\hbar}{mc}c\nabla\times [\phi_n^{*}{\bf S}\phi_{n'}]. \nonumber \\
\label{eq.current}
\end{eqnarray}
The integration in eq.~(\ref{eq.localcurrent}) is carried out around site $i$,
and $a_n(i)$ is the annihilation operator of electron with the local 
orbital expressed by the wave function $\phi_n({\bf r}-{\bf r}_i)$. 
The charge and the mass of electron are denoted
as $e$ and $m$, 
and $\hbar{\bf S}$ is the spin operator of electron.
The second term in eq.~(\ref{eq.current}), which describes the scattering
of photon, will be neglected in the following. 
The approximation made by taking account of the process only on Fe atoms
may be justified at the core-level spectra, since the core state is well 
localized at Fe sites.

The absorption experiment\cite{Kubota2004} we analyse has been carried out 
on the geometry that the photon propagates along the $a$-axis with linear 
polarization, as illustrated in Fig.~\ref{fig.setup}.
Corresponding to this situation, it is convenient to 
rewrite the interaction between the matter 
and the photon in a form,
\begin{equation}
 H_{\rm int} = -e\sum_{\bf q}\sqrt{\frac{2\pi}{V\hbar\omega_{\bf q}}}
    \sum_{i} T({\bf q},{\bf e},i) c_{\bf q}
     {\rm e}^{i \textbf{q}\cdot\textbf{r}_i} + {\rm H.c.},
\end{equation}
where the transition operator $T({\bf q},{\bf e},i)$ 
is defined as $\hbar \textbf{e} \cdot \textbf{j}(\textbf{q},i)/e$.
The explicit expression of $T({\bf q},{\bf e},i)$ for the $E$1 and $E$2
transitions are given in the following subsections.

\subsection{$E1$ transition}

The transition operator $T({\bf q},{\bf e},i)$  for the $E1$ 
transition is obtained by putting 
${\rm e}^{i{\bf q}\cdot({\bf r}-{\bf r}_j)}=1$ in eq.~(\ref{eq.localcurrent}). 
Therefore it is independent of the propagation direction of photon.
For the polarization along the z-axis, the first term in eq.~(\ref{eq.current})
is rewritten by employing the following relation
\begin{equation}
 \int \phi_n^* \frac{\partial}{\partial z}\phi_{n'} 
{\rm d}^3{\bf r} = -\frac{m}{\hbar^2}(\epsilon_n-\epsilon_{n'})
  \int \phi_n^* z\phi_{n'} {\rm d}^3{\bf r},
\end{equation}
where $\epsilon_n$ and $\epsilon_{n'}$ are energy eigenvalues corresponding
to the eigenstates $\phi_n$ and $\phi_{n'}$, respectively. 
At the $K$-edge, we assign the $4p$ states to $\phi_n$ and $1s$ state to
$\phi_{n'}$. Hence the transition operator $T^{E1}$ is expressed as 
\begin{equation}
 T^{E1}({\bf q},{\bf e},i) = 
  iB^{E1} \sum_{i\sigma}N^{E1}_{\eta}
     [p'^{\dagger}_{\eta\sigma}(i)s_{\sigma}(i)
     -s^{\dagger}_{\sigma}(i)p'_{\eta\sigma}(i)].
\label{eq.e1mat}
\end{equation}
where $i$ runs over Fe sites. Non-vanishing values
of the coefficients $N^{E1}_{\eta}$'s are given by
$N_{\eta}=1/\sqrt{3}$ for the polarization along the $\eta$ $(=x,y,z)$ axis,
independent of the propagating direction of photon.
The coefficient $B^{E1}$ is defined by
\begin{equation}
 B^{E1} = (\epsilon_{4p}-\epsilon_{1s})\int_0^{\infty}
   r^3 R_{4p}(r)R_{1s}(r){\rm d}r,
\end{equation}
where $R_{1s}(r)$ and $R_{4p}(r)$ are radial wave functions 
of the $1s$ and $4p$ states with energy $\epsilon_{1s}$ and $\epsilon_{4p}$,
respectively, in the Fe atom. The energy difference may be approximate as 
$\epsilon_{4p}-\epsilon_{1s} \sim \hbar\omega_{\bf q} = \hbar cq$.
\cite{Igarashi2008}
Within the HF approximation $B^{E1}$ is estimated as
$B^{E1} \approx 1.5\times 10^{-7} {\rm cm\cdot eV}$ 
in the 1s$^2$3d$^5$4p$^{0.001}$-configuration of an Fe atom.\cite{Cowan1981} 

\subsection{$E2$ transition}

The transition operator for the $E2$ transition is extracted from
eq. ~(\ref{eq.localcurrent}) by retaining the second term 
in the expansion ${\rm e}^{i{\bf q}\cdot({\bf r}-{\bf r}_i)} \approx 1
+i{\bf q}\cdot({\bf r}-{\bf r}_i)+ \cdots$.
Let the photon be propagating along the $y$-axis with the polarization 
parallel to the $z$-axis. Then we could derive a relation,
\begin{eqnarray}
 \int \phi_n^* y\frac{\partial}{\partial z}\phi_{n'} 
{\rm d}^3{\bf r} &=& -\frac{m}{\hbar^2}(\epsilon_n-\epsilon_{n'})
  \int \phi_n^* \frac{yz}{2}\phi_{n'} {\rm d}^3{\bf r} \nonumber \\
  &+&\frac{i}{2}\int \phi_n^* L_x\phi_{n'} {\rm d}^3{\bf r},
\label{eq.e2int}
\end{eqnarray}
where $\hbar L_x$ is the orbital angular momentum operator. 
The last term should be moved into the terms of the $M1$ transition. 
\cite{Igarashi2009}
At the $K$-edge, we assign the $3d$ states to $\phi_n$ and
the $1s$ state to $\phi_{n'}$, respectively.
Hence the transition operator $T^{E2}$ may be expressed as
\begin{equation}
 T^{E2}({\bf q},{\bf e},i) = -qB^{E2} \sum_{im\sigma}N^{E2}_{m}({\bf q})
   [d^{\dagger}_{m\sigma}(i)s_{\sigma}(i)
   -s^{\dagger}_{\sigma}(i)d_{m\sigma}(i)].
\label{eq.e2mat}
\end{equation}
When the photon is propagating along the $y$-axis,
$m$ is selectively $yz$ with $N^{E2}_{yz}({\bf q})=1/(2\sqrt{15})$
in the polarization along the $z$-axis, 
and $m$ is selectively $xy$ with $N^{E2}_{xy}({\bf q})=1/(2\sqrt{15})$
in the polarization along the $x$-axis, respectively.
Note that a relation $N_m^{E2}(-\textbf{q})=-N_m^{E2}(\textbf{q})$ holds.
The $B^{E2}$ are defined by
\begin{equation}
 B^{E2} = (\epsilon_{3d}-\epsilon_{1s})\int_0^{\infty}
   r^4 R_{3d}(r)R_{1s}(r){\rm d}r,
\end{equation}
where $R_{3d}(r)$ is radial wave function of the $3d$ state with
energy $\epsilon_{3d}$ in the Fe atom.
An evaluation within the HF approximation gives 
$B^{E2}\approx 1.7\times 10^{-16}$cm$^2\cdot$eV.\cite{Cowan1981}

\section{Absorption spectra}

Restricting the processes on Fe atoms, we sum up cross sections 
at Fe sites to obtain the absorption intensity 
$I(\omega_{\bf q},{\bf e})$.
Dividing it by the incident flux $c/V$, we have
\begin{eqnarray}
 I(\omega_{\bf q},{\bf e}) &\propto&
  \frac{4\pi^2e^2}{\hbar^2c \omega_{\bf q}}
  \sum_i \sum_f |\langle \Psi_f(i)|T({\bf q},{\bf e},i)
  |\Psi_g(i)\rangle|^2 \nonumber \\
& & \times \delta(\hbar\omega_{\bf q}+E_g-E_f),
\label{eq.intensity}
\end{eqnarray}
where $T({\bf q},{\bf e},i)= T^{E1}({\bf q},{\bf e},i) 
+ T^{E2}({\bf q},{\bf e},i)$,
and $|\Psi_g(i)\rangle$ and $|\Psi_f(i)\rangle$ represent the ground and 
the final states with energy $E_g$ and $E_f$ at site $i$, respectively.
The sum over $f$ is taken over all the excited state at Fe sites.

In the Fe pre-K-edge region, the final states are constructed by perturbation 
theory starting from the states ($|\Phi_m(d^6),\underline{1s\sigma}\rangle$)
in the $d^6$ configuration with the $1s$-core hole.
Within the second-order perturbation of $\tilde{H}^{4p-3d}$, they are given by
\begin{eqnarray}
|\Psi_f(i)\rangle &=& |\Phi_m(d^6),\underline{1s\sigma}\rangle \nonumber \\
               &+& \sum_{n{\bf k}\eta'}
     |\Phi_{n}(d^5),{\bf k}\eta'\sigma,\underline{1s\sigma}\rangle 
     \frac{1}{E_f-E'_{n{\bf k}\eta'}} \nonumber \\
  &\times&\langle\Phi_{n}(d^5),{\bf k}\eta'\sigma,\underline{1s\sigma}|
  \tilde{H}^{4p-3d}|\Phi_m(d^6),\underline{1s\sigma}\rangle, 
\label{eq.final.k}
\end{eqnarray}
where $E_f$ stands for the energy of the unperturbed state. It is defined as
\begin{equation}
 E_f = E_m(d^6) - \epsilon_{1s} - E_{\rm int}(1s-d^6),\label{eq.pre.E}
\end{equation}
where
$E_{\rm int}(1s-d^6)$ is the interaction energy
between the electron in
the $1s$ states and electrons in the $3d^6$ configuration.
In the second term of eq. (\ref{eq.final.k}), $E'_{n{\bf k}\eta'}$
is defined as
\begin{equation}
 E'_{n{\bf k}\eta'} = E_n(d^5) - \epsilon_{1s} + \epsilon_{4p}({\bf k})
                  - E_{\rm int}(1s-4pd^5) , \label{eq.main.E}
\end{equation}
where $E_{\rm int}(1s-4pd^5)$ is the interaction energy
between the electron in
the $1s$ states and electrons in the $4p3d^5$ configuration.
Symbols ${\bf k}\eta'\sigma$ and 
$\underline{1s\sigma}$ appeared in the bras and kets
indicate the presence of an electron in the $4p$ 
state $({\bf k}\eta'\sigma)$ and the absence of a $1s$-core electron with 
spin $\sigma$, respectively. Since the second term of eq. (\ref{eq.final.k})
is completely evaluated by $|\Phi_m(d^6),\underline{1s\sigma}\rangle$ and $E_f$,
the label $f$ is specified by the $m$-th eigenstates of the $d^6$ 
configuration and the core-hole spin $\sigma$.
Notice that the lowest values of eqs. (\ref{eq.pre.E})
and (\ref{eq.main.E})
correspond to the positions of the pre- and main-edges,
respectively.

The sum over ${\bf k}$ may be replaced by the integral with 
the help of the $4p$ DOS.
In our numerical treatment, the position of the pre-edge energy
is adjusted to the experimental value and the difference between the
pre- and main-edges is chosen as the minimum of $E'_{n{\bf k}\eta'}$
to be $12$ eV
higher than $E_g(d^6)-\epsilon_{1s}-E_{\rm int}(1s-d^6)$.
For simplicity, the explicit dependence on site $i$ is omitted
from the right hand side of eq.~(\ref{eq.final.k}).
From these wave-functions, we obtain the expression of transition amplitudes
at site $i$ by
\begin{equation}
 M({\bf q},{\bf e},i;f)= M^{E1}({\bf q},{\bf e},i;f) 
                       + M^{E2}({\bf q},{\bf e},i;f),
\end{equation}
with
\begin{eqnarray}
 M^{E1}({\bf q},{\bf e},i;f) &\equiv& \langle\Psi_f(i)|
  T^{E1}({\bf q},{\bf e},i)|\Psi_g(i)\rangle \nonumber\\
  &=& \sum_{n{\bf k}\eta'}
  \langle\Phi_m(d^6),\underline{1s\sigma}|\tilde{H}^{4p-3d} 
      |\Phi_n(d^5),{\bf k}\eta'\sigma,\underline{1s\sigma}\rangle \nonumber\\
  &\times& 
  \frac{1}{E_f-E'_{n{\bf k}\eta'}} 
  \langle\Phi_n(d^5),{\bf k}\eta'\sigma,\underline{1s\sigma}|
  T^{E1}({\bf q},{\bf e},i)|\Phi_g(d^5)\rangle, 
\label{eq.ME1} \\
 M^{E2}({\bf q},{\bf e},i;f) &\equiv& \langle\Psi_f(i)|
  T^{E2}({\bf q},{\bf e},i)|\Psi_g(i)\rangle \nonumber\\
  &=& \langle\Phi_m(d^6),\underline{1s\sigma}
  |T^{E2}({\bf q},{\bf e},i)
   |\Phi_g(d^5)\rangle.
\label{eq.ME2}
\end{eqnarray}
With these amplitudes, eq.~(\ref{eq.intensity}) is rewritten as
\begin{eqnarray}
 I(\omega_{\bf q},{\bf q},{\bf e}) &\propto&
 \frac{1}{\hbar\omega_{\bf q}} \sum_{i}\sum_{f} 
   |M({\bf q},{\bf e},i;f)|^2 \nonumber\\
 &\times& \frac{\Gamma/\pi}
  {[\hbar\omega_{\bf q}+E_g(d^5)-E_f]^2+\Gamma^2},
\end{eqnarray} 
where the $\delta$-function is replaced by the Lorentzian function with
the life-time broadening width of $1s$-core hole $\Gamma=0.8$ eV.

Now we examine the symmetry relation of the amplitudes.
First, let the propagating direction of photon be reversed with keeping
other conditions. 
In eq.~(\ref{eq.localcurrent}), $i{\bf q}\cdot({\bf r}-{\bf r}_i)$ is to
be replaced by $-i{\bf q}\cdot({\bf r}-{\bf r}_i)$. 
We know that $N_{\eta}^{E1}$ has no dependence on $\textbf{q}$ and
that $N_m^{E2}(-\textbf{q})$ is equal to $-N_m^{E2}(\textbf{q})$. 
Since other conditions are the same, we have the new amplitudes 
$(M^{E1})'=M^{E1}$ and $(M^{E2})'=-M^{E2}$.
Second, let the local magnetic moment at each Fe atom be reversed 
with keeping the same shifts from the center of octahedron.
The reverse of the local magnetic moment corresponds to taking 
the complex conjugate of the wave functions.
Considering eq.~(\ref{eq.ME1}) together with eq.~(\ref{eq.e1mat}),
we have $(M^{E1})'=-(M^{E1})^{*}$. 
Similarly, considering eq.~(\ref{eq.ME2}) together with eq.~(\ref{eq.e2mat}),
we have $(M^{E2})'=(M^{E2})^{*}$. 
Third, let the shifts of Fe atoms from the center of octahedron be reversed 
with keeping the same local magnetic moment, which means the reversal of
the direction of the local \emph{electric} dipole moment. 
This operation causes the reversal of
the sign of $\tilde{H}^{4p-3d}$. However,
no change is brought about to the $3d$ states 
in the $3d^5$- and $3d^6$-configurations, because the ligand field 
$\tilde{H}^{3d-3d}$ varies as $\delta^2$. 
As a result, we have the new amplitude $(M^{E1})'=-M^{E1}$ 
from eq.~(\ref{eq.e1mat}) while $(M^{E2})'=M^{E2}$. 

As already stated, the direction of the local magnetic moment could be
reversed by reversing the direction of the applied magnetic field, 
since the actual material is a ferrimagnet with slightly deviating from 
a perfect antiferromagnet. 
Let $I_{\pm}(\omega_{\bf q},{\bf q},{\bf e})$ 
be the intensity for the external magnetic field along the $\pm c$ axis.
Then, from the second symmetry relation mentioned above, we have 
the average and the difference of the intensities as
\begin{eqnarray}
 \bar{I}(\omega_{\bf q},{\bf q},{\bf e}) &\equiv& \frac{1}{2}
 \left[I_{+}(\omega_{\bf q},{\bf q},{\bf e})
  +I_{-}(\omega_{\bf q},{\bf q},{\bf e})\right]
\nonumber\\
 &\propto&\frac{1}{\hbar\omega_{\bf q}}\Biggl\{ \sum_{i}\sum_{f} 
   \Bigl{[} |M^{E1}({\bf q},{\bf e},i;f)|^2 + 
            |M^{E2}({\bf q},{\bf e},i;f)|^2 \Bigr{]} \nonumber \\
 &\times& \frac{\Gamma/\pi}
  {[\hbar\omega_{\bf q}+E_g(d^5)-E_f]^2+\Gamma^2} 
\nonumber \\
 &+& \sum_{i}|B^{E1}|^2\frac{2}{3}\sum_{\bf k} \frac{\Gamma/\pi}
  {[\hbar\omega_{\bf q}+E_g(d^5)-E'_{g{\bf k}\eta}]^2+{\Gamma}^2}
 \Biggr\}, \label{eq.intave}
\end{eqnarray}
and
\begin{eqnarray}
 \Delta I(\omega_{\bf q},{\bf q},{\bf e}) &\equiv& 
  I_{+}(\omega_{\bf q},{\bf q},{\bf e})
      -I_{-}(\omega_{\bf q},{\bf q},{\bf e}) \nonumber \\
 &\propto&\frac{2}{\hbar\omega_{\bf q}} \sum_{i}\sum_{f} 
   \Biggl\{ \left[M^{E1}({\bf q},{\bf e},i;f) \right]^{*}
   M^{E2}({\bf q},{\bf e},i;f) \nonumber\\
  &+& \left[M^{E2}({\bf q},{\bf e},i;f)\right]^{*}
     M^{E1}({\bf q},{\bf e},i;f) \Biggr\} \nonumber \\
 &\times& \frac{\Gamma/\pi}
  {[\hbar\omega_{\bf q}+E_g(d^5)-E_f]^2+\Gamma^2},
\label{eq.intdiff}
\end{eqnarray} 
respectively.
The $M^{E1}$ and 
$M^{E2}$ represent the amplitudes when the magnetic field is applied 
parallel to the $c$ axis.
For the average intensity [eq.~(\ref{eq.intave})], the last term 
describes the $K$-edge spectra due to the $E1$ transition that
the $1s$ electron is excited to the $4p$ band. 
Although its main contribution is restricted in the main $K$-edge region, 
its tail spreads over the pre-$K$-edge spectra due to the life-time width.
We see that the difference intensity [eq.~(\ref{eq.intdiff})]
is brought about from the $E1$-$E2$ interference process.
According to the symmetry relations mentioned above,
it is expected to follow\cite{Kubota2004}
%
\begin{equation}
 \Delta I(\omega_{\bf q},{\bf q},{\bf e}) \propto {\bf q} \cdot
  \sum_{i} {\bf P}_{\rm loc}(i)\times {\bf M}_{\rm loc}(i) ,
\label{eq.toroidal}
\end{equation}
where ${\bf P}_{\rm loc}(i)$ and ${\bf M}_{\rm loc}(i)$ are the electric 
and the magnetic dipole moment of Fe atom at site $i$, respectively.
Note that ${\bf P}_{\rm loc}(i)$ is proportional to
$\boldsymbol{\delta}_{i} [\equiv (0,0,\delta)]$.
Then, 
the right hand side of eq.~(\ref{eq.toroidal}) is the sum of the local
toroidal moment $\boldsymbol{\tau}(i)$ 
[$\equiv\boldsymbol{\delta}_i\times {\bf M}_{\rm loc}(i)$].
\cite{Popov1998} We have already derived the same form in the optical 
absorption spectra,\cite{Igarashi2009}
which is brought about by the $E1$-$M1$ interference process.
The spectra change their sign if one of the vectors among
${\bf q}$, ${\bf P}_{\rm loc}$, or
${\bf M}_{\rm loc}$ reverses its direction.

\begin{figure}[h]
\begin{center}\includegraphics[width=8.0cm]{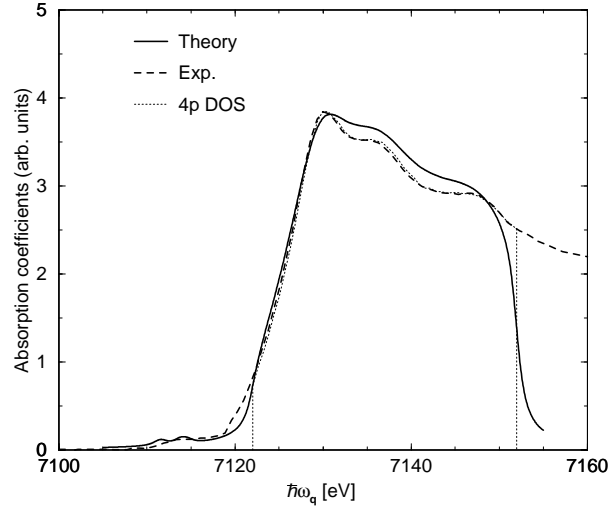}
\caption{Average intensity 
$\overline{I}(\omega_{\bf q},{\bf q},{\bf e})$ 
as a function of photon energy $\hbar\omega_{\bf q}$.
Photons propagate along the positive direction of the $a$ axis 
with polarization vector ${\bf e}$ along the $b$ axis.
The solid line represents the calculated spectra.
The broken line denotes the experimental data with the background intensity 
subtracted from the raw data given in ref.~[\citen{Jung2004}].
The dotted line represents the $4p$DOS;
both the low-energy and high-energy sides are arbitrarily cut-off.
\label{fig.average}}
\end{center}
\end{figure}

Figure \ref{fig.average} shows the calculated average intensity
$\overline{I}(\omega_{\bf q},{\bf q},{\bf e})$ 
as a function of photon energy
$\hbar\omega_{\bf q}$, in comparison with the experiment.\cite{Kubota2004}
The $1s$-core energy is adjusted such that the $K$-edge position corresponds
to the experiment.
The intensity at the $K$-edge ($\hbar\omega_{\bf q} > 7120$ eV) mainly
comes from the $E1$-$E1$ process given by the last term of 
eq.~(\ref{eq.intave}).
In the pre-$K$-edge region ($\hbar\omega_{\bf q}\sim 7110-7115$ eV),
the tail of that intensity spreads due to the life-time broadening of 
the core level. In addition, we have another contribution of
$E1$-$E1$ process through the term proportional to $|M^{E1}|^2$,
and that of the $E2$-$E2$ process through the term proportional to 
$|M^{E2}|^2$.  The latter is found larger than the former,
giving rise to a small two-peak structure with a weak polarization dependence 
(the inset in Fig.~\ref{fig.average}).
The $E1$-$E1$ process through the term proportional to $|M^{E1}|^2$
is effective only on the noncentrosymmetric situation, 
because the $4p3d^5\underline{1s}$-configuration has to mix with
the $3d^6\underline{1s}$-configuration.
Note that the $E1$-$E1$ process could also gives rise to 
the intensity in the pre-$K$-edge region through the mixing of the $4p$ state
with the $3d$ states at neighboring Fe sites. This process need not the 
noncentrosymmetric situation, and has nothing to do with the magnetoelectric
spectra. It is known that a substantial intensity of the resonant x-ray 
scattering (RXS) spectra is brought about from this process at the pre-$K$-edge 
on LaMnO$_3$,\cite{Takahashi2000} but the present analysis could not
include this process because of the cluster size.

\begin{figure}[h]
\begin{center}\includegraphics[width=8.0cm]{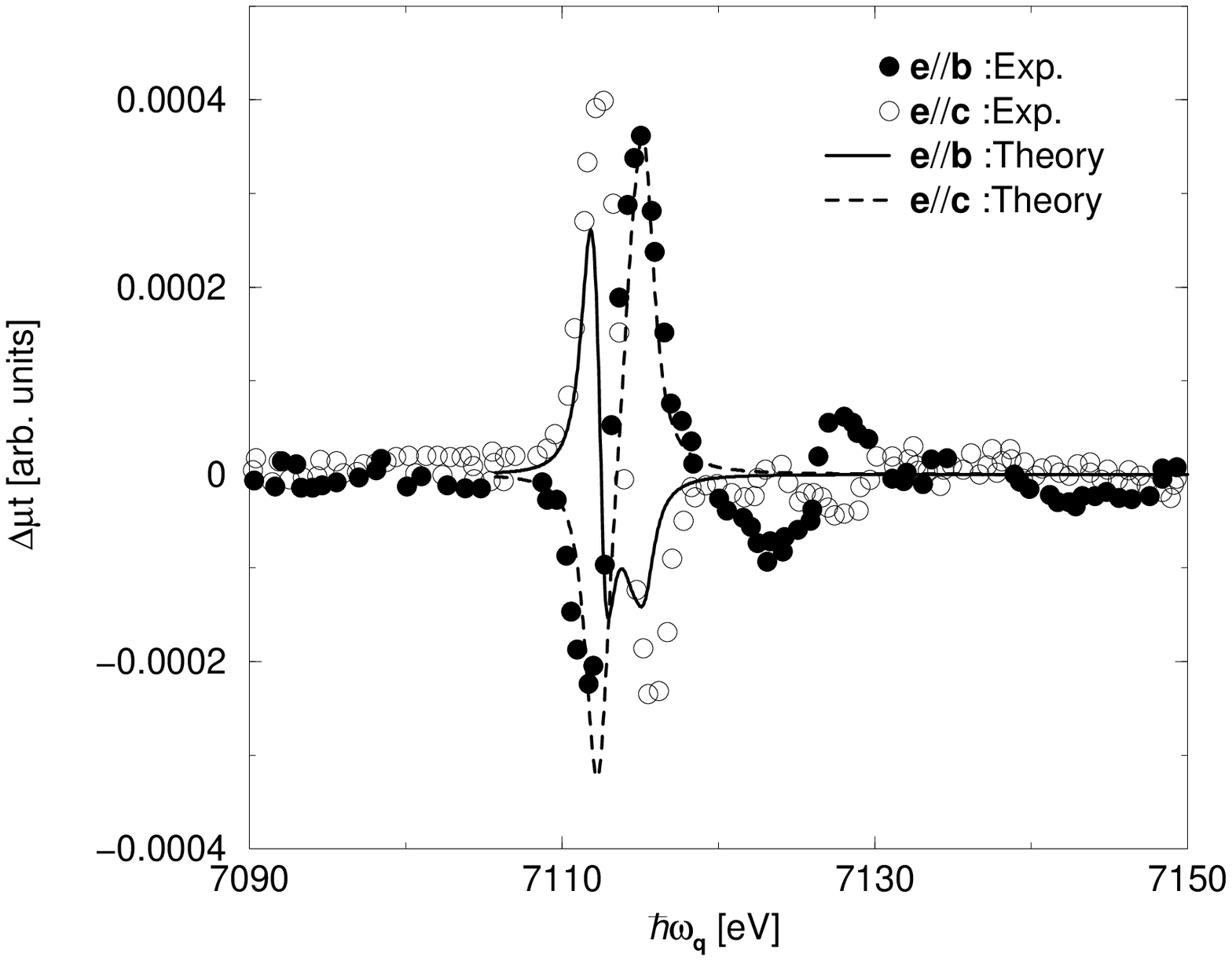}
\caption{Difference of the absorption intensities 
$\Delta I(\omega_{\bf q},{\bf q},{\bf e})$ 
as a function of photon energy $\hbar\omega_{\bf q}$
when the magnetic field is applied parallel and antiparallel to
the $c$ axis.
The solid and broken lines correspond to $\Delta I(\omega_{\bf q},{\bf q},{\bf e})$
where photons propagate along the positive direction of the $a$ axis 
with polarization vector
${\bf e}$ along the $b$ and $c$ axes, respectively.
Experimental data are taken from ref.~[\citen{Jung2004}] and
denoted as filled ($\textbf{e} \parallel b$) and open 
($\textbf{e} \parallel c$) circles, respectively.
\label{fig.difference}}
\end{center}
\end{figure}

Figure \ref{fig.difference} shows the magnetoelectric spectra 
$\Delta I(\omega_{\bf q},{\bf q},{\bf e})$ 
as a function of photon energy $\hbar\omega_{\bf q}$. 
For polarization ${\bf e}$ parallel to the $b$ axis,
the calculated spectra form a positive sharp peak and then change into
a negative double peak with increasing $\hbar\omega_{\bf q}$. 
On the other hand, for polarization ${\bf e}$ parallel 
to the $c$ axis, the calculated spectra form a negative sharp peak and 
then change into a positive sharp peak with increasing $\hbar\omega_{\bf q}$.
The corresponding experimental curves look similar, but
their signs are opposite to the calculated ones.
We do not find the origin for the opposite sign.

\section{Concluding Remarks}

We have studied the magnetoelectric effects on the x-ray absorption spectra 
in a polar ferrimagnet GaFeO$_3$. 
We have performed a microscopic calculation of the absorption spectra 
using a cluster model of FeO$_6$.
The cluster consists of an octahedron of O atoms and an Fe atom displaced
from the center of octahedron. 
We have disregarded additional small distortions of the octahedron.
We have derived an effective hybridization between the $4p$ and $3d$ states 
as well as the ligand field on the $3d$ states by modifying the Fe-O 
hybridizations due to the shifts of Fe atoms.
This leads to the mixing of the $4p3d^5$-configuration to the 
$3d^6$-configuration, and thereby to finite contributions of
the $E1$-$E2$ interference process to the magnetoelectric spectra.
We have derived the symmetry relations of the amplitudes $M^{E1}$ and
$M^{E2}$, and have discussed the directional dichroism of the spectra.
The cluster model used in the present paper is the same as the model
used in the analysis of the optical absorption spectra in GeFeO$_3$ 
and is similar to the model used in the analysis of RXS in Fe$_3$O$_4$.\cite{Igarashi2008}
We have numerically calculated the magnetoelectric spectra as a function 
of photon energy in the pre-$K$-edge region.
Although the spectral shapes are similar to the experimental curves,
their signs are opposite to the experimental ones.
The origin for the opposite sign has not been clarified yet. 
We would like to simply comment that the magnetoelectric spectra 
in the optical absorption are obtained in agreement with the experiment 
by using the same model.\cite{Igarashi2009} 

The magnetoelectric effect on RXS
has been studied experimentally\cite{Arima2005} 
and theoretically\cite{Matteo2006,Lovesey2007} in GaFeO$_3$.
We think the approach used in the present paper is
effective also to the analysis of RXS.
Closely related to these studies, the magnetoelectric effects on RXS have
also been measured,\cite{Matsubara2005,Matsubara2009} 
and theoretically analyzed\cite{Igarashi2008} in magnetite,
where A sites are tetrahedrally surrounded by oxygens with the local inversion
symmetry being broken.

\section*{Acknowledgment}
This work was partly supported by Grant-in-Aid 
for Scientific Research from the Ministry of Education, Culture, Sport, 
Science, and Technology, Japan.

\end{document}